# Thermoelectric Devices: Principles and Future Trends

**Ibrahim M Abdel-Motaleb[1,2] and Syed M. Qadri[1]**

[1]Department of Electrical Engineering, Northern Illinois University, DeKalb, IL 60115, USA

[2]Email: ibrahim@niu.edu

**Abstract**

**The principles of the thermoelectric phenomenon, including Seebeck effect, Peltier effect, and Thomson effect are discussed. The dependence of the thermoelectric devices on the figure of merit, Seebeck coefficient, electrical conductivity, and thermal conductivity are explained in details. The paper provides an overview of the different types of thermoelectric materials, explains the techniques used to grow thin films for these materials, and discusses future research and development trends for this technology.**

## 1. Introduction

Thermoelectric devices convert heat to electrical power by converting infrared photons to current. The advantages of using such devices for power generation are many. The first is that, unlike other sources of energy, thermal energy is available everywhere and anywhere. One needs only to have a differential temperature, *ΔT*, to generate electrical power. To the contrary, other means of energy sources may not be available at all times or in all places. For example, solar energy is available only during the sunny days, hydraulic energy can only be generated at water sources, and fossil fuel generators cannot operate without the availability of fuel such as oil, coal, or natural gas. The second advantage is that thermal energy is, in some cases, almost inexhaustible, especially for low energy applications. For example, thermoelectric devices can use body heat to indefinitely power some medical applications such as drug delivery, continuous diagnostics, or vital signs monitoring. The third advantage is that thermoelectric devices are reliable, maintenance free, do not use moving parts or liquid, and use direct energy conversion.

Thermoelectric effect can be used in numerous applications. Among them micro-cooling, micro-heating, powering of light sources, and recovery of wasted thermal energy from solar cells, engines, factories, or any other heat shedding systems. The attraction of using thermoelectric devices in biomedical applications can be attributed to the devices light weight, the ability of supplying unlimited power from the body heat, the safety, the stability, and the reliability.

The goal of this paper is to provide the readers with an overview about the principle of operation of thermoelectric devices and how they can be built. The paper starts with explaining the principles of thermoelectric effects. Next, the different materials used to build the devices and their growth techniques are presented. Finally, future trends are discussed to shed some light on the promising research topics.

## 2. Thermoelectric Phenomenon

### 2.1 Seebeck Effect

In 1821, Seebeck observed that: if two different or dissimilar materials are joined together and the junctions between them are held at different temperatures ($T$ and $T+\Delta T$), then a voltage difference ($\Delta V$) is developed. This voltage is found to be proportional to the temperature difference ($\Delta T$); see Figure 1. This phenomenon is called the Seebeck effect. The ratio of the voltage generated to the temperature gradient is an intrinsic property of the material, and it is called the Seebeck coefficient, S, where $S=-\Delta V/\Delta T$ [1, 2]. The temperature difference results in moving the mobile charge carriers (electrons or holes) towards the cold junction and leaving behind the oppositely charged and immobile nuclei at the hot junction. Charge movement results in the rise of a thermoelectric voltage [1-3].

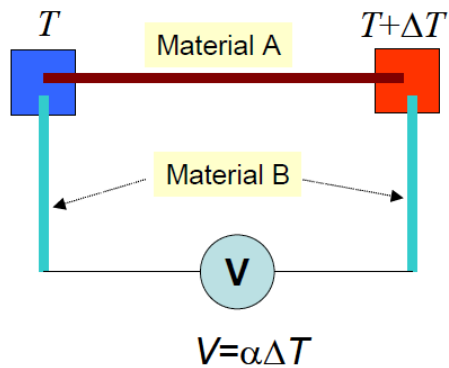

Figure 1: A diagram showing Seebeck effect. The temperature difference between the two junctions is $\Delta T$ [4].

The thermoelectric voltage ($V$) generated is given by equation (1),

$$V = (S_A - S_B)\Delta T, \tag{1}$$

where $S_A$ and $S_B$ are the Seebeck coefficients of the materials $A$ and $B$, respectively, and $\Delta T$ is the temperature difference between the junctions [1, 2].

### 2.2 Peltier Effect

Peltier effect is another important principle of the thermoelectric phenomenon. This effect explains that if an electrical current is flown through the junction of two dissimilar materials, heat will be generated or absorbed, depending on the direction of the current; see Figure 2. This effect is due to the difference in the Fermi energies of the two materials [1].

The Peltier heat, ($Q$), absorbed or rejected by the junction is given by equation (2),

$$\frac{dQ}{dT} = (\Pi_A - \Pi_B), \tag{2}$$

where $\Pi_A$ and $\Pi_B$ are called the Peltier coefficient of materials A and B, respectively, and $I$ is the current passed through the materials. The Peltier coefficient represents the amount of heat absorbed by the material when current is passed through it [3].

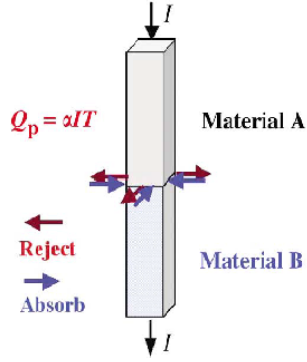

Figure 2: A diagram explaining Peltier Effect. Depending on the direction of the current, heat can be emitted or consumed in the junction. This leads to either a heater of a cooler junction [1]. In this figure α is Seebeck coefficient.

## 2.3 Thomson Effect

The Thomson effect is a combination of Seebeck and Peltier effects. This effect was discovered by William Thomson in 1854. It states that any current carrying conductor with a temperature difference between two points will either absorb or emit heat depending on the material. The heat absorbed or emitted is called the Thomson heat ($Q$) and is given by equation (3),

$$Q = \rho J^2 - \mu^* J \frac{dT}{dx},  \qquad (3)$$

where $\rho$ is the resistivity of the material and $\frac{dT}{dx}$ is the temperature gradient along the conductor, $J$ is the current density and $\mu^*$ is the Thomson coefficient. The term $\rho J^2$ is referred to as the Joule heating (irreversible) and the second term is the Thomson heating whose sign changes with the direction of the current [5].

## 3. Analysis of Thermoelectric Behavior

### 3.1 Figure of merit (*ZT*)

In 1911, Altenkirch introduced the concept of a dimensionless unit (figure of merit) to evaluate the performance of thermoelectric materials [6]. The figure of merit is defined by equation (4),

$$ZT = \frac{\sigma S^2 T}{k}, \qquad (4)$$

where $\sigma$, $S$, and $k$ are the electrical conductivity, Seebeck coefficient and thermal conductivity of the material, respectively. Therefore a good thermoelectric material should have high values for both electrical conductivity and Seebeck coefficient, but needs to have low thermal conductivity. The high electrical conductivity reduces the joule heating and increases the thermoelectric voltage generated, while the low thermal conductivity reduces of transfer of heat between the junctions.

A thermoelectric device consists mainly of thermoelectric couples comprising of a p-type and an n-type material. The thermoelectric figure of merit of such a device is given by equation (5) [1],

$$ZT = \frac{(S_p - S_n)^2 T}{\left[(\rho_n k_n)^{\frac{1}{2}} + (\rho_p k_p)^{\frac{1}{2}}\right]^2}, \tag{5}$$

where $S_p$, $\rho_p$, and $k_p$ are the Seebeck coefficient, the resistivity, and the thermal conductivity of the p-type material, respectively; and $S_n$, $\rho_n$, and $k_n$ are the Seebeck coefficient, the resistivity, and the thermal conductivity of the n-type material, respectively.

### 3.2 Seebeck coefficient or Thermo-power (*S* or *α*)

The Seebeck coefficient or the thermos-power of the material is the ratio of the induced thermoelectric voltage to the temperature difference across the material. The Seebeck coefficient of a good thermoelectric material (degenerate, highly doped semiconductor) is usually in the range of hundreds of µV/K and is given by equation (6) [3].

$$S = \frac{8\pi^2 k_B^2}{3eh^2} m^* T \left(\frac{\pi}{3n}\right)^{\frac{2}{3}}, \tag{6}$$

where $k_B$ is the Boltzmann's constant, *e* is the electronic charge, *h* is the Planck's constant, *T* is the temperature, $m^*$ is the effective mass of the carrier, and *n* is the carrier concentration. From equation (7), it can be shown that, at a given temperature, Seebeck coefficient is proportional to effective mass and the inverse of the carrier concentration, i.e.,

$$S \propto \frac{m^*}{n^{\frac{2}{3}}} \tag{7}$$

Therefore, to have a large Seebeck coefficient, the material should have high effective mass and low carrier concentration. For this reason, semiconductors have large Seebeck coefficients compared with metal.

The Seebeck coefficient for semiconductor materials, based on the band model for finely grained material, is given by equation (8) [7].

$$S = \pm \frac{k_B}{e}\left[(2 + r) + \ln\frac{2\left(2\pi m^*_{p(n)} k_B T\right)^{\frac{3}{2}}}{h^3 p(n)}\right], \tag{8}$$

where *r* is the scattering factor, which ranges from -0.9 to -0.2 for $Bi_{0.5}Sb_{1.5}Te_3$, $m^*_{p(n)}$ is the effective mass of holes (electrons) and *p(n)* is the carrier concentration. From equation 8, it can be seen that Seebeck coefficient is inversely proportional to the carrier concentration, as explained before.

### 3.3 Electrical conductivity (*σ*)

The electrical conductivity is the measure of the material's ability to conduct electric current. Electrical resistivity (*ρ*) is the reciprocal of the electrical conductivity. The unit of electrical conductivity is Siemens per meter (S.m$^{-1}$) and is given by equation (9) [1].

$$\sigma = \frac{1}{\rho} = ne\mu, \tag{9}$$

where $\mu$ is the mobility of the charge carriers, which obtained from equation (10):

$$\mu = \frac{e\tau}{m^*},  \qquad (10)$$

where $\tau$ is the mean scattering time between the collisions for the carriers.

Therefore it can be said that the electrical conductivity is directly proportional to the carrier concentration. Metals have large carrier concentrations, typically in the range of $10^{22}$ carriers per cm$^{-3}$. The electrical conductivity of the metals is typically in the order of $10^6$ ($\Omega$ cm)$^{-1}$. In case of semiconductors, thermal excitation is required for conduction to take place. Hence the conductivity is expressed for intrinsic semiconductor by [1],

$$\sigma = \sigma_0 \, exp\left(-\frac{E_G}{2k_B T}\right), \qquad (11)$$

where $\sigma_0$ is constant and $E_G$ is the energy gap for the semiconductor. Since in a semiconductor both the electrons and the holes contributes towards current conduction, the electrical conductivity can be expressed as,

$$\sigma = ne\mu_e + peu_h, \qquad (12)$$

where $n$ and $p$ are the electron and hole concentration, respectively, and $\mu_e$ and $\mu_h$ are the mobility of electrons and holes respectively. Therefore, in a semiconductor, high electrical conductivity can be achieved by having a very small energy band gap or high carrier concentration. The typical values of electrical conductivity for a semiconductor material lie in the range of $10^{-4}$ to $10^4$ ($\Omega$ cm)$^{-1}$.

### 3.4 Thermal Conductivity ($k$)

Thermal conductivity is the ability of the material to transfer heat under the effect of temperature gradient across its points. It is related to the transfer of heat either though the electron or though the vibrations of the lattice, known as phonons [1]. Thermal conductivity can be expressed as:

$$k = k_E + k_L, \qquad (13)$$

where $k_E$ and $k_L$ are the thermal conductivities due to electrons and phonons, respectively. The thermal conductivity due to electrons only is defined by equation (14) [8],

$$k_E = \frac{\pi^2 n k_B^2 T \tau}{m} = \frac{\pi^2 n^2 k_B^2 T \mu}{e}. \qquad (14)$$

The thermal conductivity due to phonon is given by equation (15) [1, 9],

$$k_L = \frac{Cvl}{3}, \qquad (15)$$

where $C$ is the phonon heat capacity per unit volume, $v$ is the average phonon velocity, and $l$ is the phonon mean free path. Since the mean free path is inversely proportional to the number of excited phonons, it can be said that thermal conductivity due to phonons increases as temperature decreases. The thermal conductivity and the electrical conductivity are related by Wiedemann Franz law.

$$k_E = L_0 \sigma T, \qquad (16)$$

where $L_0$ is a constant called Lorentz's number, and its value is $2.4 \times 10^8$ J$^2$ K$^{-2}$ C$^{-2}$. Equation (14) shows that $k_E$ is directly proportional to the mobility and the square of the concentration of the charge carriers.

Therefore, the effect of carrier mobility on the thermal conductivity is not as prominent. Thermal conductivity due to electrons can be neglected at lower temperatures since its contribution to the total thermal conductivity becomes small.

The choice of a thermoelectric material depends on the figure of merit, which depends on Seebeck coefficient and the conductivity. Table 1 compares the thermoelectric properties for metals, semiconductors and insulators. From this Table, it can be observed that metals have high electrical conductivity, but they have relatively low Seebeck coefficient. This leads to a low figure of merit. Consequently, metals are not considered the best materials for thermoelectric applications. On the other hand, insulators have high Seebeck coefficient, but they have very low electrical conductivity. This results also in a small figure of merit, which makes them unsuitable for thermoelectric applications. Contrary to metals and insulators, semiconductors have high figure of merit, because they have high electrical conductivity and relatively high Seebeck coefficient. This makes semiconductors the most suitable for thermoelectric applications.

Table 1: Average values of thermoelectric parameters of metals, semiconductors and insulators at 300K [10].

| Property | Metals | Semiconductors | Insulators |
|---|---|---|---|
| $S(\mu V K^{-1})$ | ~5 | ~200 | ~1000 |
| $\sigma(\Omega^{-1} cm^{-1})$ | ~$10^6$ | ~$10^3$ | ~$10^{-12}$ |
| $Z(K^{-1})$ | ~3 x $10^{-6}$ | ~2 x $10^{-3}$ | ~5 x $10^{-17}$ |

Figure 3 shows Seebeck coefficient, $S$, electrical conductivity, $\sigma$, the electronic thermal conductivity ($k_E$), the lattice thermal conductivity ($k_L$), and the term $S^2\sigma$, which is directly proportional to the figure of merit, $ZT$, as functions of the free-charge-carrier concentration ($n$) [10]. The figure shows that optimal thermoelectric material that has a large value of $S^2\sigma$ is located in the semiconductor region close to the metal [1, 10]. This means that the best material should be a semiconductor material with a narrow band gap, hence having high intrinsic carrier concentration.

## 4. Applications of Thermoelectric Effect

### 4.1 Thermoelectric cooling

Peltier effect is used for refrigeration, cooling of electrical components, portable coolers, cameras, climate controlled jackets, space crafts, satellites, and others. Thermoelectric devices basically consist of p-type and n-type materials connected electrically in series and thermally in parallel. Figure 4 shows the application of Peltier effect in the cooling mode [11].

When a voltage is applied at the bottom junction, free carriers move from the top junction towards the bottom, creating a temperature gradient bringing the heat from the top junction to the bottom. The efficiency of the device is directly related to the figure of merit, as explained theoretically by Ioffe [6]. The coefficient of performance ($\varphi$) for the thermoelectric cooler is given by equation (17) [1]:

$$\varphi = \frac{Q_C}{P} = \frac{(s_p-s_n)IT_C - k\Delta T - \frac{1}{2}I^2 R}{I[(s_p-s_n)\Delta T + IR]} \qquad (17)$$

where $Q_C$ is the rate of cooling, $P$ is the power consumed, $T_C$ ($T_H$) is the cold side (hot side) temperature. $\Delta T = T_H - T_C$, $I$ is the current, and $R$ is the total resistance. Thermoelectric coolers are suitable for use in micro applications because they are compact in size, require low maintenance, have no moving parts, and is reliable.

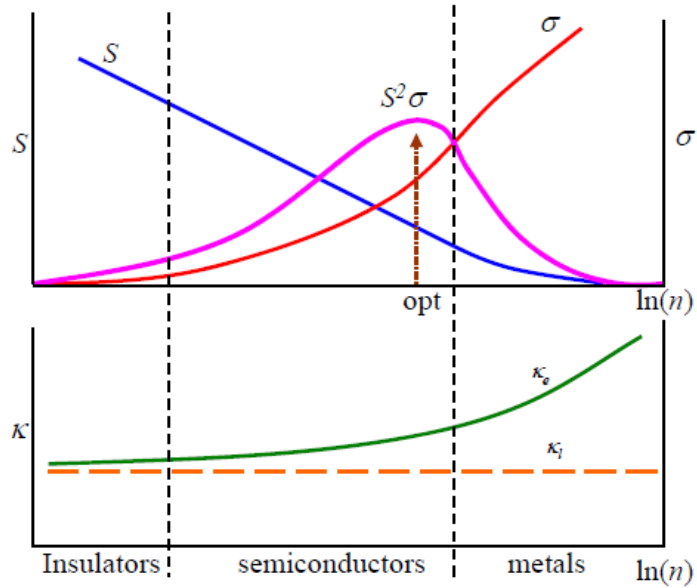

Figure 3: The relationship between Seebeck coefficient $S$, electrical conductivity $\sigma$, $S^2\sigma$ and electronic ($k_E$) and lattice ($k_L$) thermal conductivity and the free-charge-carrier concentration ($n$) of the material [10].

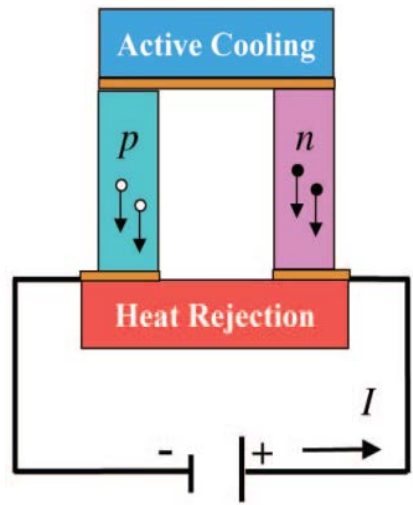

Refrigeration Mode

Figure 4: Illustration of the application of Peltier effect for refrigeration mode [11].

## 4.2 Thermoelectric power generation

Seebeck effect is used in power generation by converting the wasted-heat energy from industrial plants, vehicles, boilers, stoves, or any heat emitting system into useful power. It can also be used to salvage wasted heat from solar cells, power wrist watches, and operate space probes [12].

Figure 5 shows how Seebeck effect can be used in power generation [11]. This device maintains a temperature gradient between the source junction and the sink junction. The heat source junction maintains high temperature while the heat sink junction absorbs the heat maintaining the junction cold with respect to the heat source junction. Hence, a temperature gradient is maintained between the two junctions. This gradient makes the charge carriers diffuse from the hot side to the cold side, resulting in voltage or current flow across the junctions. The coefficient of performance, or the efficiency ($\eta$) for the thermoelectric power generator is given by the equation (8) [1].

$$\eta = \frac{P}{Q_H} = \frac{I[(s_p - s_n)\Delta T - IR]}{(s_p - s_n)IT_H + k\Delta T - \frac{1}{2}I^2 R} \tag{8}$$

Equation 8 can be rearranged and rewritten as [11]:

$$\eta = \left(\frac{T_H - T_C}{T_H}\right)\left[\frac{(1+ZT_M)^{\frac{1}{2}} - 1}{(1+ZT_M)^{\frac{1}{2}} + \left(\frac{T_C}{T_H}\right)}\right], \tag{9}$$

where $P$ is the power generated, $Q_H$ is the heat absorbed by the heat source, and $T_M$ is the average temperature. The efficiency is proportional to $(1+ZT)^{1/2}$, and it reaches Carnot efficiency if $ZT$ were to reach infinity. The efficiency of thermoelectric generators is much less than that of mechanical generator. Therefore, a great effort is directed to finding the proper materials to increase its efficiency.

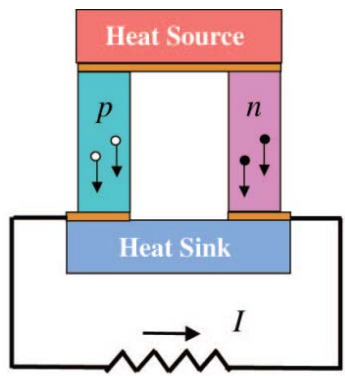

Power-Generation Mode

Figure 5: Illustration of the application of Seebeck effect for power generation [11].

## 5. Thermoelectric Materials

In 1995, Slack described the characteristics required for a material to be good thermoelectric material as a material with a narrow band gap semiconductor, high carrier mobility, and low thermal conductivity [13]. Similarly, Mahan stated that a good thermoelectric material should typically be a narrow band gap semiconductor containing high mobility carriers [14, 15].

Slack indicated that the best thermoelectric material should have the electrical properties of a crystalline material and the thermal properties of a glass material i.e. phonon glass electron crystal (PGEC) [13]. Theoretically, there is no restriction on the values of the figure of merit. However, Slack estimated that the figure of merit ($ZT$) of an optimized PGEC material is about 4 for a temperature range from 77K to 300K. For this reason, researchers are focusing on several classes of thermoelectric materials that have the potential of achieving high figure of merits, as it is discussed below.

### 5.1 Chalcogenides Materials

Chalcogenides materials contain one or more chalcogen elements such as S, Se or Te as a substantial constituent. $CsBi_4Te_6$ has been synthesized and its thermoelectric properties have been investigated by Chung [16]. When this material is doped appropriately, it exhibited a figure of merit of $ZT \sim 0.8$ at 225 K [16]. The thermoelectric properties of the two ternary tellurides ($Tl_2SnTe_5$ and $Tl_2GeTe_5$) were reported by Sharp [17]. The thermoelectric figure of merit of these materials reached a value of 0.6 at 300K and it was estimated to peak to 0.85 at 400K. PbTe compounds have exhibited a maximum figure of merit of 0.8 at 800K. However modified PbTe compounds have been found to have a $ZT$ value of more than 1 at high temperatures (around 900K) [18]. Penta tellurides have relatively high Seebeck coefficient at low temperatures, below 250K. For example, $HfTe_5$ and $ZfTe_5$ exhibits good thermoelectric properties, as reported by [19].

### 5.2 Half Heusler Alloys

Half Heusler alloys are inter metallic compounds of the general notation $M$NiSn where $M$ is a group 4 transition metals like Zr, Hf, or Ti. The potential of $TiNiSn_{1-x}Sb_x$ as a thermoelectric material has been studied by varying the doping of Sb over Sn [20]. The doping of Sb led to the maximum power factor of 1 $Wm^{-1}K^{-1}$ at room temperature for small concentration of Sb.

The thermoelectric properties of $Zr_{0.5}Hf_{0.5}NiSb_xSn_{1-x}$ were investigated by changing the doping concentration of Sb and the annealing time [21]. The minimum thermal conductivity of around 6 $Wm^{-1}K^{-1}$ was achieved in the temperature range of 50K to 150K. Electrical resistivity of 1000 μΩ-m and Seebeck coefficient of -10 μ/K was observed at 150K. The effect of partial substitution of Ni by Pd on the thermoelectric properties of ZrNiSn system was reported in [22]. $Zf_{0.5}Hf_{0.5}Ni_{0.8}Pd_{0.2}Sn_{0.99}Sb$ exhibited a figure of merit of 0.7 at 800K.

### 5.3 Skutterudites

The name of "Skutterudite" originates from the name of a small Norwegian town called Skutterud where $CoAs_3$ was extensively mined. Skutterudites are family of compounds with the general representation $MX_3$ where $M$ is a transition metal such as Co, Rh or Ir, and $X$ is pnictide elements such as P, As, or Sb. Skutterudites contain 32 atoms per unit cell, with the metal atoms on the corners of the eight ''cubes'' with six (four atom) Sb rings inside the cubes with two ''voids'' in the remaining ''cubes''

Figure of merit values of 1 - 1.3 at elevated temperatures of 700K – 900K were reported for this class of materials [23]. These high values are mainly due to the possibility of varying the lattice thermal conductivity by filling the voids within the structure with small diameter, large mass interstitials such as trivalent rare earth ions [24]. It was reported that $(Ce_yFe_{1-y})_xCo_{4-x}$ has a figure of merit of 1.4 at 1000K [1]. Yb partially filled Skutterudites and Eu doped Co based Skutterudites have exhibited a *ZT* >1 for temperatures around 600K [25, 26].

### 5.4 Clathrates

Clathrates have a cage like structure similar to that of Skutterudites. They can be classified into two types. Type 1 has the general representation of $X_2Y_6E_{46}$, while Type 2 has the general notation of $X_8Y_{16}E_{136}$. Here *X* and *Y* are alkali, alkaline earth, or rare earth metals, and *E* refers to a group 14 element, such as Si, Ge, Sn. $X_8Ga_{16}Ge_{30}$ crystals were synthesized and its properties were investigated [27]. The lattice thermal conductivities of $Eu_8Ga_{16}Ge_{30}$ and $Sr_8Ga_{16}Ge_{30}$ single crystals showed all of the characteristics of a structural glass. The thermal conductivity of $Ba_8Ga_{16}Ge_{30}$ was found to be low at room temperature with a value of ~1.3 W m$^{-1}$K$^{-1}$. High Seebeck coefficient and electrical conductivity of $Sr_8Ga_{16}Ge_{30}$ have been reported [28]. The figure of merit was found to be 0.25 at room temperature, and it is estimated to be > 1, for high temperatures. An improved value of lattice thermal conductivity (1.6 Wm$^{-1}$K$^{-1}$) under pressure (7 GPa) was reported in [29]. The total thermal conductivity was found to be 3.4 Wm$^{-1}$K$^{-1}$ and the figure of merit reached 0.75.

### 5.5 TAGS

TAGS refer to alloys of Te-Ag-Ge-Sb. Alloys of $AgSbTe_2$ have been reported with *ZT*>1 at higher temperatures for both n-type and p-type materials [30]. The p-type alloy $(GeTe)_{0.85}(AgSbTe_2)_{0.15}$ was reported to exhibit a *ZT* greater than 1.2 [31]. Seebeck coefficient for $Ag_{6.52}Sb_{6.52}Ge_{36.96}Te_{50}$ (TAGS-85) was enhanced by doping its narrow band with 1 or 2% of rare earth dysprosium (Dy) [32]. Because of the doping, the thermo-power increased from 28 μWcm$^{-1}$K$^{-2}$ to 35 μWcm$^{-1}$K$^{-2}$, and the *ZT* increased from maximum of 1.3 to more than 1.5 at 730K, making it a promising thermoelectric material. Figure 6 shows the figure of merit for n-type materials (on the left side) and p-type materials (on the right side). The shows that TAGS reaches the highest value among all materials at about 450 °C.

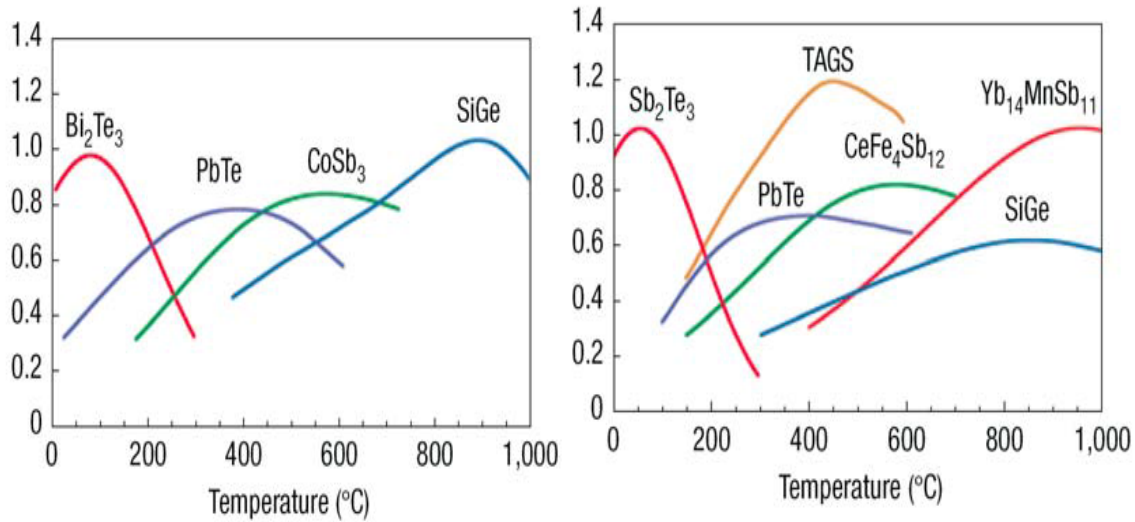

Figure 6: Figure of merit of typical thermoelectric materials (a) n-type materials (b) p-type materials [30].

## 5.6 Cobaltites

Cobaltites have high Seebeck coefficient and low thermal conductivity making them good candidates for thermoelectric materials. Typical examples of Cobaltites include $Na_xCoO_2$, $Ca_3Co_4O_9$, $Re_{1-x}Ca_xCoO_3$, and $SrCoO_{2.5}$ [4]. $Ca_3Co_4O_9$ has been grown directly on glass substrate using the PLD technique [33]. A Seebeck coefficient of 130 µV/K and resistivity of 4.3 mΩ-cm were obtained.

## 5.7 Thermoelectric Oxide Materials

Oxide materials are more rugged, easy to prepare and can be used in a wide range of operating temperatures. The properties of single-crystal $NaCo_2O_4$, which is a metallic transition-metal oxide consisting of a two-dimensional triangle lattice of Co, were measured and analyzed [34]. The in-plane resistivity was found to be 200 µΩ.cm. The Seebeck coefficient was found to be 100 mV/K, which is very larger compared to that of the metals. The large thermoelectric power and the low resistivity suggest that $NaCo_2O_4$ has a potential to be efficient thermoelectric material.

## 5.8 Bismuth telluride ($Bi_2Te_3$) and Antimony telluride ($Sb_2Te_3$)

Bismuth telluride and Antimony telluride have been widely investigated as a thermoelectric material operating at 300K. In 1954, Goldsmid attributed the excellent thermoelectric properties of $Bi_2Te_3$ to the large mean molecular mass, low melting temperature, and partial degeneracy of the conduction and valence bands [31]. It has a non-cubic tetradymite structure with space group R3m with a lattice being stacked with repeated sequence of five atom layers ($Te_1$-Bi-$Te_2$-Bi-$Te_1$) as shown in the Figure 7. The Te-Bi is held together by strong ionic-covalent bonds, whereas the Te-Te bonds between cells are weak Van der Wall forces.

Although many studies have been done on $Bi_2Te_3$, the potential of $Sb_2Te_3$ has not received the same attention until recently. $Sb_2Te_3$ has a similar structure to that of $Bi_2Te_3$. The low thermal conductivity, the high electrical conductivity, and the relatively good figure of merit ($ZT$=0.78), makes $Sb_2Te_3$ a good choice for use with $Bi_2Te_3$ to build efficient thermoelectric devices [31].

## 6. Thin Film Deposition Techniques for Thermoelectric Device

Although bulk thermoelectric materials can be used for a wide range of power generation and cooling applications, it is difficult to accommodate them for micro-scale applications. In such cases, a thin film approach is needed, where the films can be directly deposited on the substrate. Since the films are not formed under equilibrium conditions, the properties of the materials may differ from their bulk counterpart. It has been reported that using thin films instead of bulk material will result is a significant increase in the figure of merit [35]. Harman reported that the figure of merit of thin film materials is enhanced compared with bulk materials [36].

There are different methods that can be used for thin film deposition. The preferred deposition technique for a material depends on many factors, including the melting point of the material, the temperature of the deposition, and the bonding between the substrate and the material, among other factors.

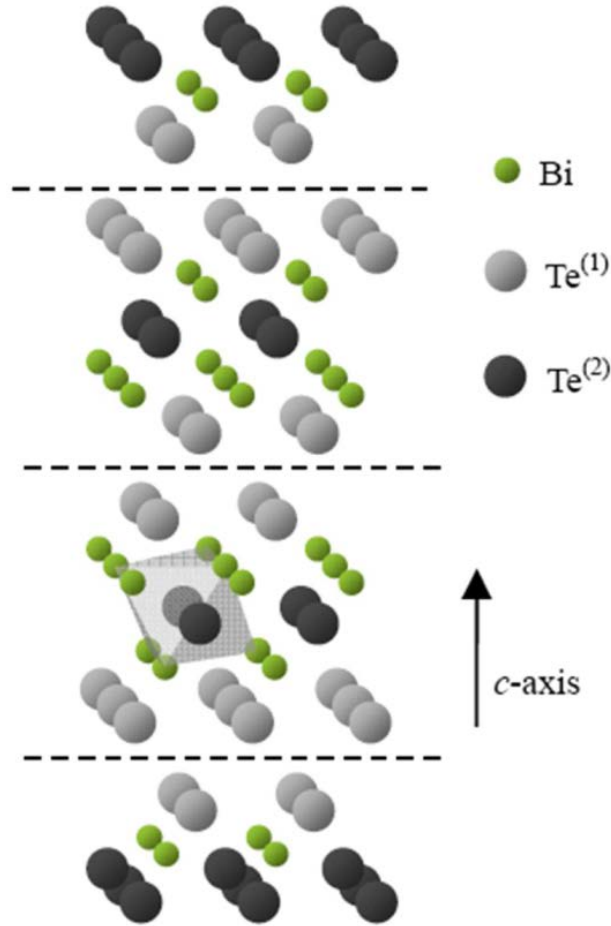

Figure 7: Atomic layers in the $Bi_2Te_3$ crystal structure. Dashed lines indicate Van der Walls gaps [31].

### 6.1 Sputtering

Sputtering utilizes high energy ions to knock target atoms to be deposited on the substrate in an inert gas atmosphere, such as Argon. RF power sources or direct current is used to ionize the Argon ions and provide them with the energy to co sputter from individual element targets or a mixture of them. After films are deposited either at high or room temperatures) they may be annealed at higher temperatures [37]. $Bi_2Te_3$ films were prepared using sputtering in which mixed powders of different compositions were used as sputtering targets. High temperature treatments resulted in the increase in the electrical conductivity and the decrease of the Seebeck coefficient, and the highest power factor was achieved when annealing temperature was 300˚C [38].

### 6.2 Molecular Beam Epitaxy (MBE)

Molecular beam epitaxy (MBE) is one of the most advanced technique used in depositing single crystals. In this technique pure materials are evaporated from individual effusion cells that are separated from the substrate by shutters or valves, while the substrate remains heated. The slow deposition rates enable the films to grow in epitaxial manner. They can be optimized to reach high level of purity by using high vacuum chambers [39]. $Bi_2Te_3$ was deposited using this technique with a Seebeck coefficient of 180 µV/K [40]. $(BiSb)_2Te_3$ was also deposited using the same technique, achieving a Seebeck coefficient of 184 µV/K and a power factor of 1.6 $mWK^{-2}m^{-1}$ [41].

### 6.3 Metal Organic Chemical Vapor Deposition (MOCVD)

Metal organic chemical vapor deposition is achieved by surface reaction of organic compounds and metal hydrides containing the required chemical elements. The film is formed by the final pyrolysis of the constituent chemicals at the surface of the substrate. The pressure applied is normally the optimum pressure required for the chemical reaction to take place. The growth temperature is usually between 300˚C to 500˚C, depending on the reaction taking place. High quality n-type and p-type materials can be prepared using this method.

$Bi_2Te_3$ and $Sb_2Te_3$ films were deposited using this technique, where Seebeck coefficient values of -210 µV/K and 110 µV/K and resistivity values of 9 µΩ-m and 3.5 µΩ-m were achieved for the two films, respectively [42]. $Sb_2Te_3$ films were also deposited at 450˚C with a maximum Seebeck coefficient of 115 µV/K [43].

### 6.4 Electrochemical Deposition (ECD)

Although electrochemical deposition provides a cheap and convenient method of preparing thin films when compared to vacuum based methods, low quality materials and inferior performance limits its application. The constituent elements are dissolved in a nitric acid solution which may also act as a chelating agent preventing the precipitation of insoluble oxides.

The n-type $Bi_2Te_3$ was deposited at a constant potential in a standard three electrode configuration [44]. $Sb_2Te_3$ films were deposited using ECD technique with power factor of 0.57 $mWK^{-2}m^{-1}$ [45]. $Bi_2Te_3$ films that electro deposited at 50 mV exhibited a Seebeck coefficient of 51.6 µV/K and a power factor of 0.71 $mWK^{-2}m^{-1}$, while the $Sb_2Te_3$ films electroplated at 20 mV exhibited a Seebeck coefficient of 52.1 µV/K and a power factor of 0.17 $mWK^{-2}m^{-1}$ [46].

### 6.5 Flash Evaporation (FE)

Flash evaporation is mainly used in deposition of thin film alloys whose constituents have different vapor pressures. In this technique a boat is maintained at sufficiently high temperature to evaporate the least volatile component of the alloy. Maintaining the critical vapor pressure and temperature of the constituents is not required, making it less complex. Unlike other powder dispensing techniques, which may require complex mechanical devices, expensive cleaning process, or is contaminative in nature, this technique requires neither complex evaporation equipment nor special precursors [47].

$(Bi_2Te_3)_{0.9}(BiSe_3)_{0.1}$ for n type material powder and $(Bi_2Te_3)_{0.25}(Sb_2Te_3)_{0.75}$ for p type material powder were evaporated using flash evaporation technique yielding figure of merit $ZT=0.21 \times 10^{-4}$ $K^{-1}$ and $ZT=0.17 \times 10^{-3}$ $K^{-1}$, respectively, at 300K [48].

## 6.6 Thermal Evaporation

Thermal evaporation is a process in where solid materials are heated inside a high vacuum chamber until they generate vapor pressure. Unlike MBE where electron beam is used to generate vapor, thermal evaporation uses thermal energy to produce vapor of the material.

Seebeck coefficient and electrical conductivity (resistivity) of p-type $Sb_2Te_3$ and n-type $Bi_2Te_3$ thin films deposited by evaporation technique were found to be 160 µV/K, $3.12 \times 10^{-3}$ Ω.cm and -200 µV/K, $1.29 \times 10^{-3}$ Ω.cm respectively [49]. High power factors of 4.9 $mWK^{-2}m^{-1}$ for $Bi_2Te_3$ and 2.8 $mWK^{-2}m^{-1}$ for $Sb_2Te_3$ have been achieved using this technique [50].

## 6.7 Pulsed Laser Deposition (PLD)

In this technique, thin films are deposited by the ablation of one or more targets. This ablation is achieved using a focused pulsed laser beam. The materials are vaporized from the target and are deposited on the substrate as thin films under high vacuum conditions in the presence of gas [51-52]. Figure 8 shows a schematic diagram of a typical PLD set up [51].

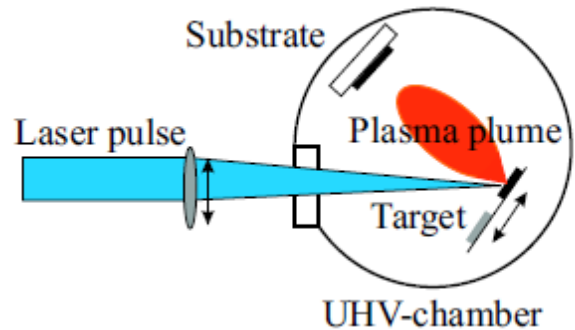

Figure 8: Schematic diagram of the deposition chamber of a typical PLD system [51].

Under high vacuum conditions, targets are struck at an angle of 45˚ by a pulsed and focused laser beam. This results in the ablation of the atoms and ions of the targets which are then deposited on the substrate. Depositions are carried out in the presence of inert gases like Argon or in the presence of Oxygen if oxide formation is required. Mostly, the substrates are attached parallel to the target surface at a target to substrate distance of 2-10 cm.

First attempt on deposition of $Bi_2Te_3$ films using PLD was reported by Dauscher in 1996 [53]. Thin films of $Bi_2Te_3$ of 60 nm thickness, with laser energy varying from 300 to 680 mJ, laser density varying from 2 to 10 $Jcm^{-2}$ and substrate temperature varying from 20˚C - 500˚C were deposited using PLD technique [54]. $Bi_2Te_3$ films have been grown on silicon and mica substrates at different temperatures ranging from 30˚C to 400˚C in a deposition chamber allowing background inert gas pressure control from $10^6$ Pa to atmospheric pressure [55]. Thin films of p-type $Bi_{0.5}Sb_{1.5}Te_3$, n-type $Bi_2Te_{2.7}Se_{0.3}$, and n-type $(Bi_2Te_3)_{90}(Sb_2Te_3)_5(Sb_2Se_3)_5$ were deposited on substrates of mica and aluminum nitride (on silicon) using pulsed laser ablation at substrate temperatures between 300 °C to 500 °C [56].

## 7. Conclusion and future trends

Thermoelectric devices will be used extensively for personal electronic applications. These applications extends from wearable items, such as watches or cloths, to biomedical applications such as drug delivery and vital signs monitoring. Current technology can provide devices for micro-thermoelectric applications, such as cooling of integrated circuit. But, to reach the full potential of these application, a cost effective mass production techniques needs to be developed, and the performance needs to be optimized. In general, mass production will extend the use of these devices to many industrial applications.

As indicated, a good thermoelectric material should have high electrical conductivity but low thermal conductivity [13]. Electrical conductivity increases with the increase of the carrier concentration and and/or the carrier mobility. Therefore, improvement can be achieved by developing material structure or fabrication process that increases the mobility or the carrier concentration, or both. One way to achieve this goal is to use super-lattice, quantum well structures, where the mobility can be enhanced and consequently Seebeck effect can be increased [57]. These structures benefit from quantum confinement which enhances the figure of merit and this can be extended to nanowire and quantum dots.

Another way of increasing the carrier concentration is by bombarding a multilayer thermoelectric material with Si ions at different doses. This forms quantum structures in the multilayers, which leads to improving the efficiency as claimed by reference [58]. The order of the thermoelectric layers is as follows: 6 multilayers of $SiO_2/SiO_2 + Ge$, single layer of Ge, single layer of Sb + Ge, single layer of Si, single layer of Si + Ge, and 10 multilayers of Ge/Si + Ge. Reference [58] claims that the increase of the electronic density of states in the mini-band of the quantum structures is due to the ion beam bombardment, which increases the electrical conductivity and Seebeck coefficient. It should be noted that quantum structures can also help reducing the material thermal conductivity.

The problem with increasing conductivity is that it increases thermal conductivity too. The increase of the thermal conductivity reduces Seebeck effect, which reduces the device efficiency, as explained before. The logical solution is to decouple electrical conductivity from thermal conductivity. In other words, make thermal conductivity independent of electronic conductivity. One way to do that is to use phonon glass electron crystal (PGEC) [13]. This material structure can be created by alloying, or site substitution, with isoelectric elements such as $Bi_2Te_3$ with $Sb_2Te_3$ or $Bi_2Se_3$. This forms a p-type $Bi_xSb_{1-x}Te_3$ and n-$Bi_2Se_yTe_{1-y}$. Another technique is to introduce disorder into a complex lattice structure [59]. For example adding defect atoms in the void space of the material structure would result in a PGEC material. This is because the disorder is located in the void; hence, it will not affect conductivity, since transport takes place outside the void. On the other hand, defect atoms will suppress the thermal conductivity due to phone scattering.

Decoupling the electric and thermal conductivities can also be achieved by using the spin Seebeck effect (SSE) [60]. In this effect the thermally generated spin current is converted into current via spin-orbit interaction in conducting materials in a close proximity to the magnets [60]. Figure 9 shows a schematic for the SSE based devices using ferromagnetic insulator adjacent to a conducting layer. Charges are generated in the low thermal conductivity ferromagnetic material but conducted using a highly conductive material. In this figure, part (a) shows a schematic illustrations of the fundamental element of the thermoelectric device based on the conventional Seebeck effect, while part (b) shows a module structure for the same device. Similarly, parts (c) and (d) shows the fundamental element and module structure of a device based on the longitudinal spin Seebeck effect (LSSE). Part (e) is schematic illustrations of the LSSE-device structures being developed for future thermoelectric applications. In this figure, as stated by [60], ∇T, $E_{SE(ISHE)}$, M, and $j_S$ refers to the temperature gradient, electric field generated by the Seebeck effect [inverse spin Hall effect (ISHE)], magnetization vector, and spatial direction of the thermally generated spin current, respectively. is the length of the metallic film of the LSSE device along the x (y) direction $l_{x(y)}$, and $l_z$ is the thickness of the metallic film [60].

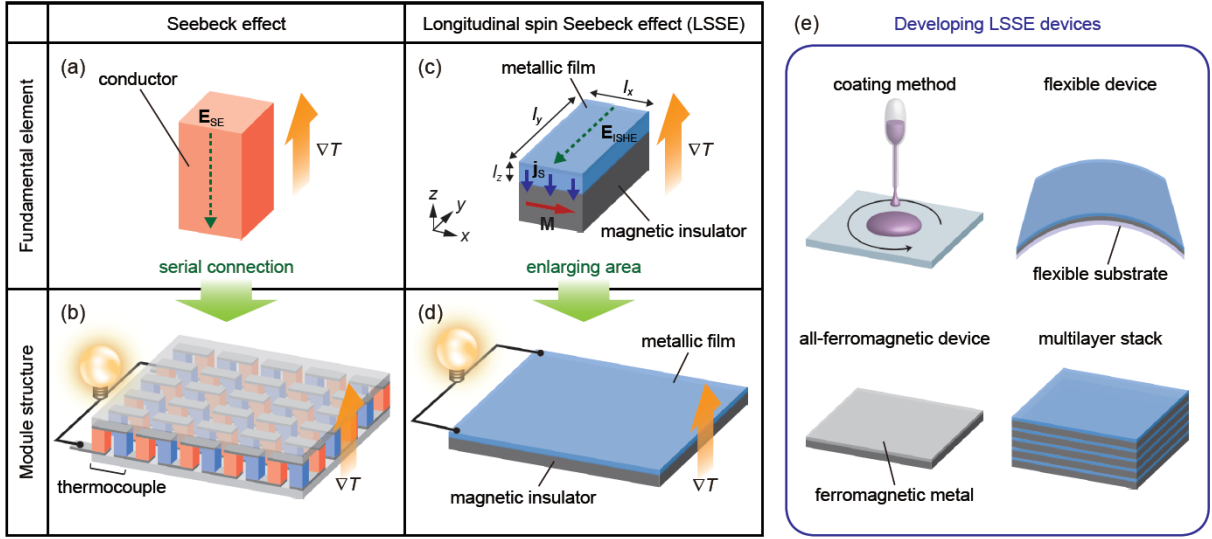

Figure 9: Device structure for Longitudinal Spin Seebeck Effect (LSSE).

Lately, Topological Insulator (TI) has been found to provide high figure of merit for thermoelectric devices [61]. A combination of $Bi_2Te_3$ and $Sb_2T_3$ was found to be a topological insulator. These materials has distinct transport behaviors. The surface Dirac fermions dominate magneto-electric transport while the thermoelectric effect is mainly determined by the bulk states. Using this material, it was possible to achieve a figure of merit ZT up to 1.86, which considerably higher than conventional thermoelectric device [62].

In the USA, there is a plan to use thermoelectric devices in automobile industry. The Department of Energy is funding Ford and GM companies to develop thermoelectric devices system for vehicle Heating, Ventilation, and Air Condition (HVAC) [63]. The success of this project will widen thermoelectric devices use in the industrial and electronic applications.

In conclusion, thermoelectric devices have a great potential in both personal and industrial electronics. However, a more efficient materials need to be developed and a mass production techniques needs to be adopted to reach this potential.

**References**

[1] T. M. Tritt, '*Thermoelectric Materials: Principles, Structure, Properties, and Applications*', Encyclopedia of Materials: Science and Technology, pp: 1-11, 2002, ISBN: 0-08-043152-6.

[2] http://www.dea.icai.upco.es/romano/sp4/tpares_all.pdf
[3] Andreas Willfahart, Screen Printed Thermoelectric Devices," Linköpings Universitet, SE-601 74 Norrköping, 2014
[4] Jin-Cheng Zheng, '*Recent advances on thermoelectric materials*', Front. Phys. China, 2008, 3(3): 269-279.
[5] Xingcun Colin Tong, '*Advanced Materials for Thermal Management of Electronic Packaging*', Volume 30, 2011, ISBN: 978-1-4419-7758-8.
[6] G.S. Nolas, J. Sharp, J. Goldsmid, '*Thermoelectrics: Basic Principles and New Materials Developments*', Springer, 2001.
[7] Kazuhiro Hasezaki, Takashi Hamachiyo, Maki Ashida, Takashi Ueda and Yasutoshi Noda, '*Thermoelectric Properties and Scattering Factors of Finely Grained Bi2Te3-Related Materials Prepared by Mechanical Alloying;* Materials Transactions, Vol. 51, No. 5 (2010) pp. 863 to 867.
[8] Kittel C., '*Introduction to Solid State Physics*', Wiley, 2005.
[9] Ibach, H., Luth, H., '*Solid-State Physics: An Introduction to Principles of Materials Science*', Springer, 2009, ISBN 978-3-540-93803-3.
[10] C. Wood, '*Materials for thermoelectric energy conversion*', Reports on Progress in Physics, 1988, Volume 51, Number 4, Pg 459
[11] Terry M. Tritt and M.A. Subramanian, '*Materials, Phenomena, and Applications: A Bird's Eye View*', MRS Bulletin, Volume 31, March 2006.
[12] http://en.wikipedia.org/wiki/Thermoelectric_generator.
[13] G.A. Slack, '*New materials and performance limits for thermoelectric cooling*', in CRC Handbook of Thermoelectrics, ed. by D.M. Rowe, CRC Press, Boca Raton, FL, 1995, p. 407.
[14] G. D. Mahan, J. O. Sofo, '*The best thermoelectric*', Proc. Natl. Acad. Sci. USA, Vol. 93, pp. 7436-7439, July 1996, Applied Physical Sciences.
[15] G. Mahan, B. Sales and J. Sharp, Phys. Today 50 (3), 42 (1997).
[16] Duck-Young Chung et al., '*CsBi$_4$Te$_6$: A High-Performance Thermoelectric Material for Low-Temperature Applications*', Science 287, 1024 (2000).
[17] Jeff W. Sharp, Brian C. Sales, David G. Mandrus, and Bryan C. Chakoumakos, '*Thermoelectric properties of Tl$_2$SnTe$_5$ and Tl$_2$GeTe$_5$*', Applied Physics Letters, Volume 74, Number 25, 21 June 1999.
[18] Z.H. Dughaish, '*Lead telluride as a thermoelectric material for thermoelectric power generation*', Physica B 322 (2002), 205–223.
[19] Littleton IV R T, Tritt T M, Korzenski M, Ketchum D, Kolis J W, 2001, '*Effect of Sb substitution on the thermoelectric properties of the Group IV pentatelluride materials $M_{1-x}Y_xTe_5$ (M¼Hf, Zr and Ti)*', Phys. Rev. B. Rapid Comm. 64, 121104–7.
[20] S. Bhattacharya, Terry M. Tritt, Y. Xia, V. Ponnambalam, S. J. Poon, and N. Thadhani, '*Grain structure effects on the lattice thermal conductivity of Ti-based half-Heusler Alloys*', Applied Physics Letters 81, 43 (2002).
[21] C. Uher, J. Yang, S. Hu, D. T. Morelli and G. P. Meisner, '*Transport properties of pure and doped MNiSn, M= (Zr, Hf)*', Physical Review B, Volume 59, Number 13, 1 April 1999.
[22] Q. Shen, L. Chen, T. Goto, T. Hirai, J. Yang, G. P. Meisner, and C. Uher,, '*Effects of partial substitution of Ni by Pd on the thermoelectric properties of ZrNiSn-based half Heusler compounds*', Applied Physics Letters 79, 4165 (2001).
[23] X. Shi, W. Zhang, L. D. Chen, J. Yang and C. Uher, '*Theoretical study of the filling fraction limits for impurities in CoSb$_3$*', Physical Review B 75, 235208, (2007).

[24] J. Yang, G. P. Meisner, D. T. Morelli and C. Uher, '*Iron valence in skutterudites: Transport and magnetic properties of Co$_{1-x}$Fe$_x$Sb$_3$*', Physical Review B, Volume 63, 014410, 2000.

[25] G. S. Nolas, M. Kaeser, R. T. Littleton IV, and T. M. Tritt, '*High figure of merit in partially filled ytterbium skutterudite materials*', Applied Physics Letters 77, 1855 (2000).

[26] G. A. Lamberton Jr., S. Bhattacharya, R. T. Littleton IV, M. A. Kaeser, R. H. Tedstrom, T. M. Tritt, J. Yang, and G. S. Nolas, '*High figure of merit in Eu-filled CoSb$_3$-based skutterudites*', Applied Physics Letters 80, 598 (2002).

[27] B. C. Sales, B. C. Chakoumakos, R. Jin, J. R. Thompson and D. Mandrus, '*Structural, magnetic, thermal, and transport properties of X$_8$Ga$_{16}$Ge$_{30}$ (X=Eu, Sr, Ba) single crystals*', Physical Review B, Volume 63, 245113, 2001.

[28] G. S. Nolas, J. L. Cohn, G. A. Slack and S. B. Schujman, '*Semiconducting Ge clathrates: Promising candidates for thermoelectric applications*', Applied Physics Letters, Volume 73, Number 2, 13 July 1998.

[29] J. F. Meng, N. V. Chandra Shekar, J. V. Badding, G. S. Nolas, '*Threefold enhancement of the thermoelectric figure of merit for pressure tuned Sr$_8$Ga$_{16}$Ge$_{30}$*', Journal of Applied Physics, Volume 89, Number 3, 1 February 2001.

[30] G. Jeffrey Snyder and Eric S. Toberer, '*Complex Thermoelectric Materials*', Nature Materials, Volume 7, February 2008.

[31] Skrabek E. A. and Trimmer D. S. in '*CRC Handbook of Thermoelectrics*', ed. Rowe, D. M., 267–275, CRC, Boca Raton, 1995.

[32] Levin E. M., Bud'ko, S. L., and Schmidt-Rohr K., '*Enhancement of Thermopower of TAGS-85 High-Performance Thermoelectric Material by Doping with the Rare Earth Dy*', US Department of Energy Publications, Paper 334, 2012.

[33] Y. F. Hu, E. Sutter, W. D. Si, and Qiang Li, '*Thermoelectric properties and microstructure of c -axis-oriented Ca$_3$Co$_4$O$_9$ thin films on glass substrates*', Applied Physics Letters 87, 171912 (2005).

[34] I. Terasaki, Y. Sasago and K. Uchinokura, '*Large thermoelectric power in NaCo$_2$O$_4$ single crystals*', Physical Review B, Third Series, Volume 56, Number 20, 15 November 1997.

[35] Rama Venkatasubramanian, Edward Silvola, Thomas Colpitts and Brooks O'Quinn, '*Thermoelectric devices with room-temperature figures of merit*', Nature, Volume 13, 11 October 2001.

[36] T. C. Harman, et al., '*Quantum Dot Superlattice Thermoelectric Materials and Devices*', Science 297, 2229 (2002).

[37] Harald Böttner, Joachim Nurnus, Alexander Gavrikov, Gerd Kühner, Martin Jägle, Christa Künzel, Dietmar Eberhard, Gerd Plescher, Axel Schubert and Karl-Heinz Schlereth, '*New Thermoelectric Components Using Microsystem Technologies*', Journal of Microelectromechanical Systems, Vol. 13, No. 3, June 2004.

[38] Huang, H., Luan, W. and Tu, S., '*Influence of annealing on thermoelectric properties of bismuth telluride films grown via radio frequency magnetron sputtering*', Thin Solid Films, 517, 2009, pp. 3731–3734.

[39] John R. Arthur, '*Molecular beam epitaxy,*' ELSEVIER-Surface Science, Vol. 500, 2002, pp 189-217.

[40] Charles E., Groubert E. and Boyer A., '*Structural and electrical properties of bismuth telluride films grown by the molecular beam technique*', Journal of Materials Science Letters 7, 6 (1988), 575-577.

[41] S. Cho1, Y. Kim, J. B. Ketterson, '*Structural and thermoelectric properties in $(Sb_{1-x}Bi_x)_2Te_3$ thin films*', Appl. Phys. A 79, 1729–1731 (2004).

[42] A. Giani, A. Boulouz, F. Pascal-Delannoy, A. Foucaran, E. Charles, A. Boyer, '*Growth of $Bi_2Te_3$ and $Sb_2Te_3$ thin films by MOCVD*', Materials Science and Engineering: B 09/1999, 64(1):19–24.

[43] A. Giani, A. Boulouz, F. Pascal-Delannoy, A. Foucaran, A. Boyer, '*Electrical and thermoelectrical properties of $Sb_2Te_3$ prepared by the metal-organic chemical vapor deposition technique*', Journal of Materials Science Letters, 18 (1999), 541-543.

[44] J. R. Lim, G. J. Snyder, C. K. Huang, J. A. Herman, M. A. Ryan and J. P. Fleurial, '*Thermoelectric Microdevice Fabrication Process and Evaluation at the Jet Propulsion Laboratory (JPL)*', Jet Propulsion Laboratory, California Institute of Technology.

[45] Lim S.K., Kim M.Y. and Oh, T.S., '*Thermoelectric properties of the bismuth-antimony-telluride and the antimony-telluride films processed by electrodeposition for micro-device applications*', Thin Solid Films 517, 14 (2009), 4199-4203.

[46] Min-Young Kimi and Tae-Sung Oh, '*Electrodeposition and Thermoelectric Characteristics of $Bi_2Te_3$ and $Sb_2Te_3$ Films for Thermopile Sensor Applications*', Journal of Electronic Materials, Vol. 38, No. 7, 2009.

[47] M. Hemanadhan, Ch. Bapanayya and S. C. Agarwal, '*Simple flash evaporator for making thin films of compounds*', J. Vac. Sci. Technol. A, Vol. 28, No. 4, Jul/Aug 2010.

[48] A. Foucaran, A. Sackda, A. Giani, F. Pascal-Delannoy, A. Boyer, '*Flash evaporated layers of $(Bi_2Te_3–Bi_2Se_3)_{(n)}$ and $(Bi_2Te_3–Bi_2Se_3)_{(p)}$*', Mater. Sci. Eng. B, 52 (1998), 154–161.

[49] Helin Zou, D. M. Rowe and S. G. K. Williams, '*Peltier effect in a co-evaporated $(Sb_2Te_3)_{(p)}$-$(Bi_2Te_3)_{(n)}$ thin film thermocouple*', Thin Solid Film, Volume 408, Issues 1–2, 3 April 2002, Pages 270–274.

[50] Carmo J. P., Goncalves L. M., Wolffenbuttel R.F., and Correia J. H., '*A planar thermoelectric power generator for integration in wearable microsystems*', Sensors and Actuators A: Physical 161, 1 (2010), 199-204.

[51] Hans-Ulrich Krebs, Martin Weisheit, Jorg Faupel, Erik Suske, Thorsten Scharf, Christian Fuhse, Michael Stormer, Kai Sturm, Michael Seibt, Harald Kijewski, Dorit Nelke, Elena Panchenko and Michael Buback, '*Pulsed Laser Deposition (PLD) - a Versatile Thin Film Technique*', Advances in Solid State Physics, Volume 43, 2003, pp 505-518.

[52] D. B. Chrisey and G. K. Hubler, '*Pulsed Laser Deposition of Thin Film*', John Wiley & Sons, Inc., New York (1994).

[53] A. Dauscher, A. Thomy and H. Scherrer, '*Pulsed laser deposition of $Bi_2Te_3$ thin film*', Thin Solid Films, Volume 280, Issues 1–2, July 1996, Pages 61–66.

[54] R. Zeipl, M. Pavelka, M. Jelinek, J. Chval, P. Lostak, K. Zd'ansky, J. Vanis, S. Karamazov, S. Vackova and J. Walachova, '*Some properties of very thin $Bi_2Te_3$ layers prepared by laser ablation*', Phys. Stat. Sol. (c) 0, No. 3, 867–871 (2003).

[55] A. Bailini, F. Donati, M. Zamboni, V. Russo, M. Passoni, C.S. Casari, A. Li Bassi, C.E. Bottani, '*Pulsed laser deposition of $Bi_2Te_3$ thermoelectric films*', Applied Surface Science, 254 (2007), 1249–1254.

[56] Raghuveer S. Makala, K. Jagannadham and B. C. Sales, '*Pulsed laser deposition of $Bi_2Te_3$-based thermoelectric thin films*', Journal of Applied Physics 94, 3907 (2003).

[57] Yehea Ismail Ahmed Al-Askalany, '*Thermoelectric Devices Cooling and Power Generation*' https://arxiv.org/ftp/arxiv/papers/1403/1403.3836.pdf.

[58] Satilmis, Budak; Zhigang, Xiao; Barry, Johnson; Jordan Cole; Mebougna Drabo; Ashley Tramble; and Chauncy, Casselberry; '*Highly-Efficient Advanced Thermoelectric Devices from Different Multilayer Thin Films*,' American Journal of Engineering and Applied Sciences 2016, 9 (2): 356.363.
[59] Snyder G J and Toberer E S 2008 *Nature Mater.* 7 105.
[60] Ken-ichi Uchida, Hiroto Adachi, Takashi Kikkawa, Akihiro Kirihara, Masahiko Ishida, Shinichi Yorozu, Sadamichi Maekawa, and Eiji Saitoh, 'ArXiv:1604.00477v2, [Cond-Mat-Mtrl-Sci], 21 May, 2016. https://arxiv.org/pdf/1604.00477.pdf
[61] Jinsong Zhang, Xiao Feng, Yong Xu, Minghua Guo, Zuocheng Zhang, Yunbo Ou, Yang Feng, Kang Li, Haijun Zhang, Lili Wang, Xi Chen, Zhongxue Gan, Shou-Cheng Zhang, Ke He, Xucun Ma, Qi-Kun Xue, and Yayu Wang, '*Disentangling the magnetoelectric and thermoelectric transport in topological insulator thin films*' Phys. Rev. 2015, *B* 91 075431.
[62] Kim S I, Lee K H, Mun H A, Kim H S, Hwang S W, Roh J W, Yang D J, Shin W H, Li X S, Lee Y Hm G J Snyder, and SW Kim, '*Thermoelectrics. Dense dislocation arrays embedded in grain boundaries for high-performance bulk thermoelectrics* ' 2015 Science, 2015 April 3, 348 (6230) p.109-14
[63] Carl Maronde, Clay Maranville Edward Gundlach, '*Automotive thermoelectric HVAC development and demonstration project*' National Energy Technology Laboratory United States Department of Energy, May 2016. http://www.energy.ca.gov/2016publications/CEC-500-2016-017/CEC-500-2016-017.pdf.